\pdfoutput=1
\documentclass{svmult}

\usepackage[T1]{fontenc}
\usepackage{times}			

\usepackage{makeidx}         
\usepackage{multicol}        
\usepackage[bottom]{footmisc}	

\usepackage{amsmath} 
\usepackage{amssymb} 
\usepackage[mathscr]{eucal} 
\usepackage[final]{graphicx}        
\usepackage{color}
\usepackage{hyperref}

\DeclareMathOperator{\Aut}{Aut}
\newcommand{\bigast}{\mbox{\Large $*$}} 
\newcommand{\bsy}[1]{\boldsymbol{#1}} 
\newcommand{\dd}{\ensuremath{\check{d}}}  
\newcommand{\diag}{\mbox{\rm diag}}
\newcommand{\gam}[1]{\mbox{$\gamma_{\!\ssst#1}$}} 
\newcommand{\order}{\mbox{\rm order}}
 
\newcommand{\ssst}{\scriptscriptstyle}
\newcommand{\suppint}{\mbox{\rm supp\_int}}	
\newcommand{\supprad}{\mbox{\rm supp\_rad}}	
\newcommand{\zinv}{\mbox{$z^{-1}$}}

\smartqed		

\begin{document}

\title*{On the Group-Theoretic Structure of Lifted Filter Banks}
\author{Christopher M.\ Brislawn\\
(Final corrections, updated for arXiv.org, 9/29/13)}
\authorrunning{C.M. Brislawn} 	
\institute{Christopher M.\ Brislawn \at Los Alamos National Laboratory, Los Alamos, NM, 87545, \email{brislawn@lanl.gov}
}
%
%

\maketitle

\abstract{
The polyphase-with-advance matrix representations of whole-sample symmetric (WS) unimodular  filter banks form a multiplicative matrix Laurent polynomial group.  Elements of this group can always be factored into lifting matrices with half-sample symmetric (HS) off-diagonal lifting filters; such linear phase lifting factorizations are specified in the ISO/IEC JPEG~2000 image coding standard.  Half-sample symmetric unimodular filter banks do not form a group, but such filter banks can be \emph{partially} factored into a cascade of whole-sample \emph{antisymmetric} (WA) lifting matrices starting from a concentric, equal-length HS base filter bank.  An algebraic framework called a \emph{group lifting structure} has been introduced to formalize the group-theoretic aspects of matrix lifting factorizations.  Despite their pronounced differences, it has been shown that the group lifting structures for both the WS and HS classes satisfy a \emph{polyphase order-increasing property}  that implies uniqueness (``modulo rescaling'') of irreducible group lifting factorizations in both group lifting structures. These unique factorization results can in turn be used to characterize  the group-theoretic structure of the  groups generated by the WS and HS group lifting structures.
}

\abstract*{}


\section{Introduction}\label{sec:Intro}
Lifting~\cite{Sweldens96,Sweldens:98:SIAM-lifting-scheme,DaubSwel98} is a general technique for factoring the  polyphase matrix representation of a  perfect reconstruction multirate filter bank into elementary matrices over the Laurent polynomials.  As one might expect of a technique as universal as elementary matrix factorization, lifting has proven extremely useful for both theoretical investigations  and practical   applications. For instance, lifting forms the basis for  specifying discrete wavelet transforms in the ISO/IEC JPEG~2000 standards~\cite{ISO_15444_1,ISO_15444_2}.  

In addition to providing a completely general mathematical framework for standardizing discrete wavelet transforms, lifting also provides a cascade structure  for  \emph{reversible} filter banks---nonlinear implementations of linear filter banks that furnish bit-perfect invertibility in fixed-precision arithmetic~\cite{BruekersEnden:92:networks-perfect-inversion,SaidPearl93,ZandiAllenEtal::Compression-with-reversible,CaldDaubSwelYeo:ACHA-98:ints-to-ints}.  Reversibility allows digital communications  systems to realize the  efficiency and scalability of subband coding  while also providing the option of lossless transmission, a key feature that made lifting a particularly attractive choice for the JPEG~2000 standard.

The author became acquainted with lifting  while serving on the JPEG~2000 standard, and he was struck by the  group-theoretic flavor of the subject.  After completing his standards committee work, he began studying  the lifting structure of two-channel linear phase FIR filter banks in depth, leading to the publications outlined in the present paper.  In spite of its universality, lifting is not particularly well-suited for analyzing \emph{paraunitary} filter banks because, as discussed in~\cite[Section~IV]{Bris:10:GLS-I}, lifting matrices are never paraunitary.  This means  lifting factorization takes place \emph{outside} of the paraunitary group, whereas we shall show that lifting factorization can be defined to take place entirely  \emph{within} the  group of whole-sample symmetric (WS, or odd-length linear phase)  filter banks by decomposing WS filter banks into linear phase lifting steps.  This allows us to prove both existence and (rather surprisingly) uniqueness of ``irreducible''  WS group lifting factorizations.  One consequence of this unique factorization theory is that we can characterize the group-theoretic structure of the unimodular WS filter bank group up to isomorphism using standard group-theoretic constructs.

Besides WS filter banks, there is also a class of half-sample symmetric (HS, or even-length linear phase) filter banks.    The differences between the group-theoretic structure of WS and HS filter banks are striking.  For instance, HS filter banks do \emph{not} form a matrix group, but linear phase ``partial'' lifting factorizations partition the class of unimodular HS filter banks into \emph{cosets} of a particular matrix group generated by whole-sample \emph{antisymmetric} lifting filters.  The complete group-theoretic classification of unimodular HS filter banks is still incomplete as of this writing but comprises an extremely active area of research by the author.

The present paper is an expository overview of recent research 
\cite{BrisWohl06,Bris:10:GLS-I,Bris:10b:GLS-II,Brislawn:13:Group-theoretic-structure}.  It is targeted at a mathematical audience that has at least a passing familiarity with elementary group theory and with the connections between wavelet transforms and multirate filter banks.

\subsection{Perfect Reconstruction Filter Banks}\label{sec:Intro:PR}	
\begin{figure}[tb]
\sidecaption
		\includegraphics{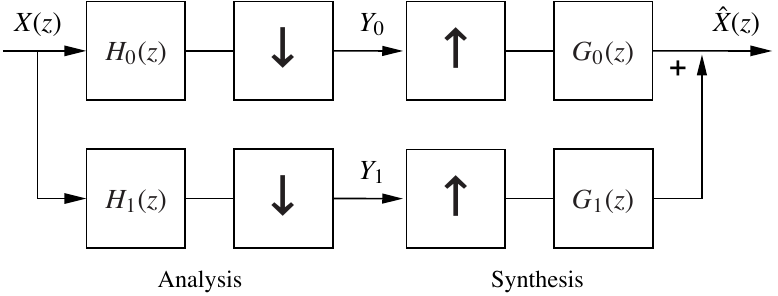}
		\caption{Two-channel perfect reconstruction multirate filter bank.}
		\label{fig:FB2chan}
\end{figure}
This paper studies two-channel multirate digital filter 
banks of the form shown in Figure~\ref{fig:FB2chan}~\cite{CroRab83,Daub92,Vaid93,VettKov95,StrNgu96,Mallat99}.
We  only consider systems in which both the analysis filters
$\left\{H_0(z),\, H_1(z)\right\}$ and the synthesis filters $\left\{G_0(z),\, G_1(z)\right\}$
are  linear translation-invariant (or time-invariant) finite impulse response (FIR) filters.  A system like that  in Figure~\ref{fig:FB2chan} is called a {\em  perfect reconstruction multirate
filter bank\/} (frequently abbreviated to just ``filter bank'' in this paper) if it is a linear translation-invariant system with a transfer function satisfying
\begin{equation}
\label{PRMFB}
\frac{\hat{X}(z)}{X(z)} = az^{-d}
\end{equation}
for some integer $d\in\mathbb{Z}$ and some constant $a\neq 0$.

FIR filters are written in the transform domain as  Laurent polynomials,
\[ F(z) \equiv \sum_{n=a}^b f(n)\,z^{-n}\in \mathbb{C}\left[z,z^{-1}\right], \]
%
%
with impulse response $f(n)$.  The {\em support interval\/} of an FIR filter, denoted
\begin{equation}\label{supp_int}
\suppint(F)\equiv \suppint(f)\equiv [a,b]\subset\mathbb{Z}, 
\end{equation}
is the smallest closed interval of integers containing the support of the filter's impulse response or, equivalently, the largest closed interval for which $f(a)\neq 0$ and \mbox{$f(b)\neq 0$.}
If $ \suppint(f)=[a,b]$ then the {\em order\/} of the filter is  
\begin{equation}\label{order}
\order(F) \equiv b-a.
\end{equation}
%

\subsection{The Polyphase-with-Advance Representation}\label{sec:Intro:Poly}
It is more efficient to compute the decimated output of a filter bank like the one  in Figure~\ref{fig:FB2chan} by splitting the signal into even- and odd-indexed subsequences,
\begin{equation}\label{xn_poly}
x_i(n) \equiv x(2n+i),\; i=0,\,1;\quad X(z) = X_0(z^2) + \zinv X_1(z^2).
\end{equation}
The {\em polyphase vector form\/} of a discrete-time signal is defined to be
\begin{equation}\label{xn_vector}
\bsy{x}(n) \equiv \left[ \begin{array}{l}
            x_0(n)\\
            x_1(n)
        \end{array} \right];\quad
\bsy{X}(z) \equiv  \left[ \begin{array}{l}
            X_0(z)\\
            X_1(z)
        \end{array} \right].
\end{equation}
The \emph{analysis polyphase-with-advance representation} of a  filter~\cite[equation~(9)]{BrisWohl06} is 
\[ f_j(n) \equiv f(2n-j),\; j=0,\,1;\quad F(z)=F_0(z^2) + zF_1(z^2). \]
Its \emph{analysis polyphase vector representation}  is
\begin{eqnarray}
\bsy{F}(z)  &\equiv& 
        \left[ \begin{array}{l}
        F_0(z)\\
        F_1(z)
        \end{array}\right]
=  \sum_{n=c}^{d}\bsy{f}(n)\,z^{-n},\label{poly_fltr}\\
\bsy{f}(n) & \equiv & 
        \left[ \begin{array}{l}
        f_0(n)\\
        f_1(n)
        \end{array}\right]
\quad\mbox{with $\bsy{f}(c),\,\bsy{f}(d)\neq\bsy{0}$.}\label{poly_fltr_impulse}
\end{eqnarray}

Since we generally work with analysis   filter bank representations,  ``polyphase'' will mean ``analysis polyphase-with-advance.'' The polyphase filter~(\ref{poly_fltr}), (\ref{poly_fltr_impulse}) has the \emph{polyphase support interval}
\begin{equation}\label{poly_vector_supp_int}
\suppint(\bsy{f}) \equiv [c,d] ,
\end{equation}
which differs from the scalar support interval~(\ref{supp_int}) for  the same filter.
The \emph{polyphase order} of~(\ref{poly_fltr}) is
\begin{equation}\label{poly_filt_order}
\mbox{order}(\bsy{F}) \equiv d-c\;.
\end{equation}

These definitions generalize for FIR filter banks, $\{H_0(z),\,H_1(z)\}$.  Decompose each filter $H_i(z)$ into its polyphase vector representation $\bsy{H}_i(z)$ as in~(\ref{poly_fltr}) and form the \emph{polyphase matrix}
\begin{eqnarray}
\mathbf{H}(z) & \equiv  & 
        \left[ \begin{array}{l}
        \bsy{H}_0^T(z)\\
        \bsy{H}_1^T(z)
        \end{array}\right]
= \sum_{n=c}^d \mathbf{h}(n)\,z^{-n},\label{poly_matrix_def}\\
\mathbf{h}(n) & \equiv & 
        \left[ \begin{array}{l}
        \bsy{h}_0^T(n)\\
        \bsy{h}_1^T(n)
        \end{array}\right]
\quad\mbox{with $\mathbf{h}(c),\,\mathbf{h}(d)\neq \mathbf{0}$.}\label{poly_matrix_impulse_def}
\end{eqnarray}
Bold italics denote column vectors and bold roman (upright) fonts denote matrices.

The polyphase support interval of the filter bank in~(\ref{poly_matrix_def}), (\ref{poly_matrix_impulse_def}) is defined to be
\begin{equation}
\suppint(\mathbf{h}) \equiv [c,d] \label{FB_suppint} ,
\end{equation}
and the polyphase order  is defined to be
\begin{equation}\label{FB_order}
\mbox{order}(\mathbf{H}) \equiv d-c.
\end{equation}
With this notation, the output of the analysis  bank  in Figure~\ref{fig:FB2chan} can be written
\[  \bsy{Y}(z) = \mathbf{H}(z)\bsy{X}(z).  \]
An analogous synthesis polyphase matrix representation, $\mathbf{G}(z)$, can be defined for the synthesis filter bank $\left\{G_0(z),\,G_1(z)\right\}$; see~\cite[Section~II-A]{BrisWohl06}.
\begin{figure}[tb]
\sidecaption
		\includegraphics{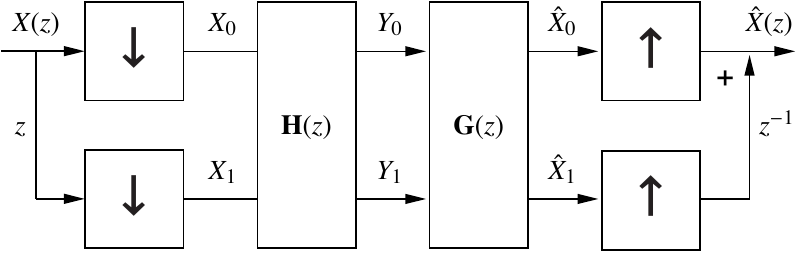}
		\caption{The polyphase-with-advance representation of a two-channel multirate filter bank.}
		\label{fig:DS_poly}
\end{figure}

The block diagram for this matrix-vector filter bank representation, which we call the \emph{polyphase-with-advance representation}~\cite{BrisWohl06}, is shown in Figure~\ref{fig:DS_poly}.  The polyphase representation transforms the non-translation-invariant analysis bank of Figure~\ref{fig:FB2chan} into a demultiplex operation, $x(k)\mapsto \bsy{x}(n)$, followed by a linear translation-invariant operator acting on vector-valued signals.  The polyphase representation therefore reduces the study of multirate filter banks  to the study of invertible transfer matrices over the Laurent polynomials.  

Since Laurent \emph{mono}mials are units, invertibility of $\mathbf{H}(z)$ over $\mathbb{C}[z,z^{-1}]$ is equivalent to
\begin{equation}\label{FIR_PR}
|\mathbf{H}(z)| \equiv \det\mathbf{H}(z)  = \check{a}z^{-\dd};\quad \check{a}\neq 0,\ \dd\in\mathbb{Z}.
\end{equation}
$\dd$ is called the \emph{determinantal delay} of $\mathbf{H}(z)$ and $\check{a}$ is called the \emph{determinantal amplitude}.
A filter bank satisfying (\ref{FIR_PR}) is called an \emph{{FIR} perfect reconstruction (PR) filter bank}~\cite{Vaid93}.
It was noted in~\cite[Theorem~1]{BrisWohl06} that the family $\mathscr{F}$ of all  FIR PR filter banks forms a nonabelian   matrix group,   called the \emph{FIR filter bank group}.  The \emph{unimodular group}, $\mathscr{N}$, is the normal subgroup of  $\mathscr{F}$ consisting of all matrices of determinant~1, 
\begin{equation}\label{DS_FIR_PR}
|\mathbf{H}(z)| = 1\,.
\end{equation}
The unimodular group can also be regarded as \mbox{$SL(2,\, \mathbb{C}\left[z,z^{-1}\right])$}.

\subsection{Linear Phase Filter Banks}\label{sec:Intro:Linear}
 It is easily shown~\cite[eqn.~(20)]{BrisWohl06} that a discrete-time signal is symmetric about one of its samples, $x(i_0)$, if and only if its polyphase vector representation (\ref{xn_vector}) satisfies
\begin{equation}\label{bfXz_symmetry}
\bsy{X}(z^{-1})=z^{i_0}\mathbf{\Lambda}(z)\bsy{X}(z),\quad\mbox{where}\quad
\mathbf{\Lambda}(z)\equiv\diag(1,z^{-1}).
\end{equation}
We say a signal satisfying (\ref{bfXz_symmetry}) is \emph{whole-sample symmetric} (WS) about $i_0\in\mathbb{Z}$.  Similarly, a discrete-time signal is \emph{half-sample symmetric} (HS) about an odd multiple of 1/2 (indexed by $i_0\in\mathbb{Z}/2$) if and only if
\begin{equation}\label{bfXz_mirror}
\bsy{X}(z^{-1})=z^{(2i_0-1)/2}\mathbf{J}\bsy{X}(z),\quad\mbox{where}\quad
\mathbf{J}\equiv 
\left[
	\begin{array}{cc}
	0 & 1 \\
	1 & 0
	\end{array}
\right] .
\end{equation}
Analogous characterizations of whole- and half-sample \emph{antisymmetry} (abbreviated WA and HA, respectively) are obtained by putting minus signs in~(\ref{bfXz_symmetry}) and~(\ref{bfXz_mirror}).  Real-valued discrete-time signals (or filters) possessing any of these symmetry properties are called \emph{linear phase} signals (filters).

\begin{figure}[tb]
\sidecaption
		\includegraphics{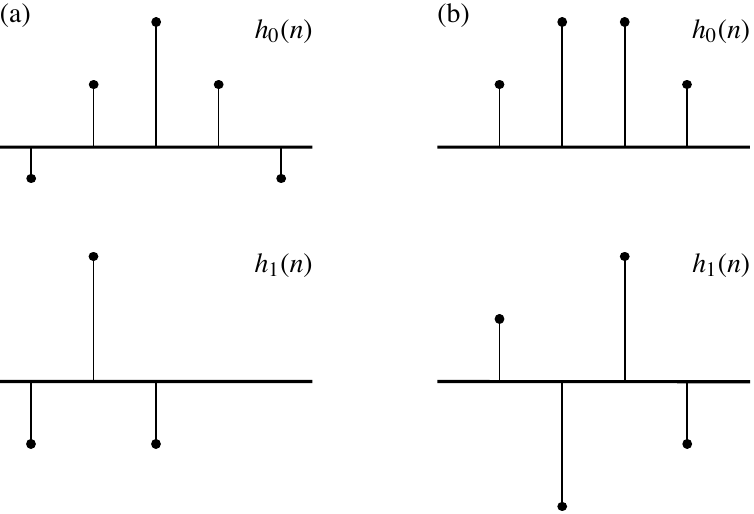}
		\caption{(a) Whole-sample symmetric filter bank. (b) Half-sample symmetric filter bank.}
		\label{fig:WS-HS}
\end{figure}
It was proven in~\cite{NguVaid:89:2-channel-PR-LP} that the only \emph{nontrivial} classes  (classes with at least one nontrivial real degree of freedom) of two-channel FIR PR linear phase filter banks  are the  whole-  and  half-sample symmetric classes shown in Figure~\ref{fig:WS-HS}.  Arbitrary combinations of symmetry  are not necessarily compatible with invertibility; e.g., if both filters have odd lengths then both must be symmetric (WS).  In an  even-length filter bank, one filter must be symmetric  (HS) while the other must be antisymmetric (HA).  It was also proven in~\cite{NguVaid:89:2-channel-PR-LP} that the sum of the impulse response lengths must be a multiple of 4, so  it is  possible for HS (but not WS) filter banks to have filters of \emph{equal lengths}, as shown in  Figure~\ref{fig:WS-HS}.  

Linear phase properties of filter banks are also straightforward to characterize in the polyphase domain~\cite[Section~III]{BrisWohl06}.  The \emph{group delay}~\cite{OppSchBuck98} of a linear phase FIR filter is equal to the midpoint (or axis of symmetry) of the filter's impulse response.  Let $d_i$ denote the group delay of $h_i$ for $i=0,1$.  
\begin{lemma}[\cite{BrisWohl06}, Lemma~2]\label{lemma_WS}
A real-coefficient FIR transfer matrix $\mathbf{H}(z)$ is a WS analysis filter bank with group 
delays $d_0$ and $d_1$ if and only if
\begin{equation}\label{WS_anal_sym}
\mathbf{H}(z^{-1}) = \diag(z^{d_0},z^{d_1})\mathbf{H}(z)\mathbf{\Lambda}(z^{-1}).
\end{equation}
\end{lemma}

If $\mathbf{H}(z)$  satisfies (\ref{FIR_PR}) then the  \emph{delay-minimized WS filter bank} normalization
\begin{equation}\label{delay-minimized_WS}
d_0 = 0,\; d_1 = -1
\end{equation}
ensures that the determinantal delay, $\dd=(d_0+d_1+1)/2$, is zero and~(\ref{WS_anal_sym}) becomes
\begin{equation}\label{WS_intertwining}
\mathbf{H}(z^{-1}) = \mathbf{\Lambda}(z)\mathbf{H}(z)\mathbf{\Lambda}(z^{-1}).
\end{equation}

The analogous delay-minimized HS filter bank normalization is 
\begin{equation}\label{delay-minimized_HS}
d_0 = -1/2 = d_1.
\end{equation}
Both filters  have the same axis of symmetry, as  in Figure~\ref{fig:WS-HS}(b); we call such  filter banks \emph{concentric}.   Delay-minimized HS filter banks are characterized by the relation
\begin{equation}\label{HS_anal_mirror_DS}
\mathbf{H}(z^{-1}) = \mathbf{L}\mathbf{H}(z)\mathbf{J}
\quad\mbox{where}\quad 
\mathbf{L} \equiv \diag(1,\,-1).
\end{equation}

We now see a striking difference between the algebraic properties of WS and HS filter banks.  Since 
\mbox{$\mathbf{\Lambda}(z^{-1})=\mathbf{\Lambda}^{-1}(z)$,} (\ref{WS_intertwining}) says that $\mathbf{\Lambda}(z)$ \emph{intertwines} $\mathbf{H}(z)$ and $\mathbf{H}(z^{-1})$, so the set of all filter banks satisfying (\ref{WS_intertwining}) (i.e., the set of all delay-minimized WS filter banks) forms a multiplicative group.  In sharp contrast, filter banks satisfying (\ref{HS_anal_mirror_DS}) do \emph{not} form a group.
\begin{definition}[\cite{Bris:10:GLS-I}, Definition~8]\label{defn:uni_WS_group}
The \emph{unimodular WS  group}, $\mathscr{W}$,  is the group of all real FIR transfer matrices that satisfy both~(\ref{DS_FIR_PR}) and~(\ref{WS_intertwining}).
\end{definition}
\begin{definition}[\cite{Bris:10:GLS-I}, Definition~9]\label{defn:uni_HS_class}
The \emph{unimodular HS  class}, $\mathfrak{H}$,  is the set of all real FIR transfer matrices that satisfy both~(\ref{DS_FIR_PR}) and~(\ref{HS_anal_mirror_DS}).
\end{definition}
%

\section{Lifting Factorization of Linear Phase Filter Banks}\label{sec:Lift}		
We now define lifting and apply it to linear phase filter banks, focusing on the problem of factoring linear phase filter banks into linear phase lifting steps.

\subsection{Lifting Factorizations}\label{sec:Lift:Lift}	
Daubechies and Sweldens~\cite{DaubSwel98} used the Euclidean algorithm for $\mathbb{C}[z,z^{-1}]$ to prove that any unimodular FIR transfer matrix can be decomposed into a {\em lifting factorization} (or  \emph{lifting cascade}) of the form
\begin{equation}\label{anal_lift_cascade}
\mathbf{H}(z) = \mathbf{D}_K\,\mathbf{S}_{N-1}(z)\cdots\mathbf{S}_1(z)\,\mathbf{S}_0(z)\;.
\end{equation}
\begin{figure}[tb]
\sidecaption
		\includegraphics{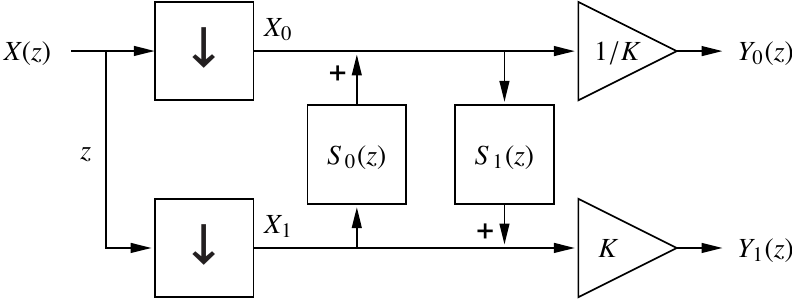}
		\caption{Two-step lifting representation of a unimodular filter bank.}
		\label{fig:irr_anal_lift}
\end{figure}
The diagonal matrix $\mathbf{D}_K\equiv\diag(1/K,\,K)$ is a \emph{unimodular gain-scaling matrix} with \emph{scaling factor} $K\neq 0$.  The lifting matrices $\mathbf{S}_i(z)$ are upper- or lower-triangular with ones on the diagonal and a  lifting filter, $S_i(z)$, in  the  off-diagonal position. 

In the factorization corresponding to  Figure~\ref{fig:irr_anal_lift},  the lifting matrix for the step  ${S}_0(z)$ (which is  a lowpass update) is upper-triangular,  and the matrix  for the second step (a highpass update) is lower-triangular.
For example, the Haar filter bank
\begin{equation}\label{Haar_filters}
H_0(z) = (z+1)/2,\quad H_1(z) = z-1,
\end{equation}
has a unimodular polyphase representation with two different lifting factorizations,
\begin{eqnarray}
\mathbf{H}_{haar}(z) \equiv 
        \left[ \begin{array}{cc}
                1/2 & 1/2 \\
                -1 & 1
        \end{array}\right]
&=& 
        \left[ \begin{array}{cc}
                1/2  & 0 \\
                0    & 2
        \end{array}\right]
        \left[ \begin{array}{cc}
                1  & 0 \\
                -1/2  & 1
        \end{array}\right]
        \left[ \begin{array}{cc}
                1  & 1 \\
                0  & 1
        \end{array}\right]\label{alt_Haar_lifting_steps}\\
&=& 
        \left[ \begin{array}{cc}
                1  & 1/2 \\
                0  & 1
        \end{array}\right]
        \left[ \begin{array}{cc}
                1  & 0 \\
                -1  & 1
        \end{array}\right]\label{Haar} .
\end{eqnarray}
Factorization (\ref{alt_Haar_lifting_steps}) fits the ladder structure of Figure~\ref{fig:irr_anal_lift} with $S_0(z)=1$, $S_1(z)=-1/2$, and $K=2$.  Factorization~(\ref{Haar}), on the other hand, begins with a \emph{highpass} lifting update and does not require a gain-scaling operation.
\begin{definition}[\cite{ISO_15444_2}, Annex~G]\label{defn:update_char}
The {\em update characteristic} of a lifting step (or lifting matrix) is a binary flag, $m=0$ or $1$, indicating which polyphase channel is being updated by the lifting step.  
\end{definition}
For instance, the update characteristic, $m_0$, of the first lifting step in Figure~\ref{fig:irr_anal_lift}  is  ``lowpass,'' coded with a zero ($m_0=0$), while the update characteristic of the second  step is ``highpass'' ($m_1=1$).  The update characteristic $m_i$ is defined similarly for each  matrix $\mathbf{S}_i(z)$ in a lifting cascade~(\ref{anal_lift_cascade}). 

Next, we generalize (\ref{anal_lift_cascade}) slightly to accommodate factorizations that lift one filter bank to another.  A \emph{partially factored lifting cascade},
\begin{equation}\label{lift_from_B}
\mathbf{H}(z) = \mathbf{D}_K\,\mathbf{S}_{N-1}(z) \cdots\mathbf{S}_0(z)\, \mathbf{B}(z),
\end{equation}
is an expansion relative to some \emph{base} filter bank, $\mathbf{B}(z)$, with scalar filters $B_0(z)$ and $B_1(z)$.  We sometimes write such factorizations in recursive form:
\begin{eqnarray}
\mathbf{H}(z) &=&  \mathbf{D}_K\,\mathbf{E}^{(N-1)}(z),\nonumber\\
\mathbf{E}^{(n)}(z) &=& \mathbf{S}_n(z)\,\mathbf{E}^{(n-1)}(z),\quad 0\leq n< N,\label{recursive_cascade}\\
\mathbf{E}^{(-1)}(z)  &\equiv&  \mathbf{B}(z).\nonumber
\end{eqnarray}
%

\subsection{Whole-Sample Symmetric Filter Banks}\label{sec:Lift:WS}	
The fact that delay-minimized WS filter banks form a group makes it easy to characterize the lifting matrices that lift one  delay-minimized WS filter bank to another,
\begin{equation}\label{WS_lift}
\mathbf{F}(z) = \mathbf{S}(z)\,\mathbf{H}(z).
\end{equation}
\begin{lemma}[\cite{BrisWohl06}, Lemma~8]\label{lemma_HS_steps}
A lifting matrix, $\mathbf{S}(z)$, lifts a  filter bank satisfying~(\ref{WS_intertwining}) to another  filter bank satisfying~(\ref{WS_intertwining}) if and only if $\mathbf{S}(z)$ also satisfies~(\ref{WS_intertwining}).
An upper-triangular lifting matrix satisfies~(\ref{WS_intertwining}) if and only if  its lifting filter is {\em half-}sample 
symmetric about $1/2$. A lower-triangular lifting matrix satisfies~(\ref{WS_intertwining}) if and only if its lifting filter is HS about $-1/2$.
\end{lemma}

Note that HS lifting \emph{filters} with  appropriate group delays  form lifting \emph{matrices} that are  WS filter banks.  It is easy to show that the lifting filters symmetric about $1/2$ form an additive group, $\mathscr{P}_0$, of Laurent polynomials and that the upper-triangular lifting matrices with lifting filters in $\mathscr{P}_0$  form a multiplicative group, $\mathscr{U}$.  Similarly, the lifting filters symmetric about $-1/2$ form an additive group, $\mathscr{P}_1$, and the lower-triangular lifting matrices with lifting filters in $\mathscr{P}_1$  form a multiplicative group, $\mathscr{L}$.

Given Lemma~\ref{lemma_HS_steps}, it is natural to ask  whether every filter bank in $\mathscr{W}$ has a lifting factorization of the form (\ref{anal_lift_cascade}) in which every lifting matrix $\mathbf{S}_i(z)$ satisfies~(\ref{WS_intertwining}).  The answer is yes, and the  proof is a constructive, order-reducing recursion that does not rely on the Euclidean algorithm.
\begin{theorem}[\cite{BrisWohl06}, Theorem~9]\label{thm_WS_factorization}
A unimodular  filter bank, $\mathbf{H}(z)$,  satisfies the delay-minimized WS condition~(\ref{WS_intertwining}) if and only if it can be factored as 
\begin{equation}\label{WS_lift_cascade}
\mathbf{H}(z) = \mathbf{D}_K\,\mathbf{S}_{N-1}(z) \cdots \mathbf{S}_1(z)\,\mathbf{S}_0(z),
\end{equation}
where each lifting matrix, $\mathbf{S}_i(z)$, satisfies~(\ref{WS_intertwining}).
\end{theorem}
We refer to such decompositions  as \emph{WS group lifting factorizations.}  This is the form of lifting factorizations specified in~\cite[Annex~G]{ISO_15444_2} for user-defined WS filter banks.  

Definition~\ref{defn:uni_WS_group} of the unimodular WS group, $\mathscr{W}$, is independent of  lifting, but we need lifting to define \emph{reversible}  WS filter banks.  Let $\mathscr{U}_r$ and $\mathscr{L}_r$ be the subgroups of $\mathscr{U}$ and $\mathscr{L}$ with  matrices whose lifting filters have \emph{dyadic} coefficients of the form $k\cdot 2^n,\, k,n\in\mathbb{Z}$.  Since gain-scaling operations are not generally invertible in fixed-precision arithmetic, gain scaling  is not used in reversible implementations.  
\begin{definition}[\cite{Bris:10:GLS-I}, Example~3]\label{defn:Wr}
The group $\mathscr{W}_r$ of reversible unimodular WS filter banks is defined to be the group of all transfer matrices $\mathbf{H}(z)$ generated by lifting factorizations (\ref{WS_lift_cascade}) where $\mathbf{S}_i(z)\in\mathscr{U}_r\cup\mathscr{L}_r$ and $\mathbf{D}_K=\mathbf{I}$.
\end{definition}
%

\subsection{Half-Sample Symmetric Filter Banks}\label{sec:Lift:HS}	
Lifting factorization of HS filter banks is harder (i.e., more interesting) than lifting factorization of WS filter banks, in  part ``because'' HS filter banks do not form a group.  For instance, the characterization in Lemma~\ref{lemma_HS_steps} of lifting matrices that lift one WS filter bank to another is equally valid for \emph{left} lifts, as in~(\ref{WS_lift}), and \emph{right} lifts in which $\mathbf{S}(z)$ acts on the right.  This  fails badly for HS filter banks.
\begin{theorem}[\cite{BrisWohl06}, Theorem~12]\label{thm_no_HS_right_lifting}
Suppose that $\mathbf{H}(z)$  is an HS filter bank satisfying the concentric delay-minimized condition~(\ref{HS_anal_mirror_DS}).  If $\mathbf{F}(z)$ is right-lifted from  $\mathbf{H}(z)$,
\[ \mathbf{F}(z) = \mathbf{H}(z)\,\mathbf{S}(z), \]
then $\mathbf{F}(z)$ can only satisfy~(\ref{HS_anal_mirror_DS}) if $\mathbf{S}(z)=\mathbf{I}$ and $\mathbf{F}(z)=\mathbf{H}(z)$.
\end{theorem}

Fortunately, half-sample symmetry can be preserved by left-lifting operations.
\begin{lemma}[\cite{BrisWohl06}, Lemma~10]\label{lemma_WA_steps}
If either $\mathbf{H}(z)$ or $\mathbf{F}(z)$ in~(\ref{WS_lift}) is an HS filter bank satisfying the concentric delay-minimized condition~(\ref{HS_anal_mirror_DS}), then the other filter bank  also satisfies~(\ref{HS_anal_mirror_DS}) if and only if $\mathbf{S}(z)$ satisfies
\begin{equation}\label{WA_lifting_filter}
\mathbf{S}(z^{-1}) = \mathbf{L}\,\mathbf{S}(z)\,\mathbf{L}  = \mathbf{S}^{-1}(z),
\end{equation}
which says that the lifting filter  is whole-sample antisymmetric (WA) about 0.
\end{lemma}

WA lifting filters form an additive group, $\mathscr{P}_a$, and the upper-triangular (resp., lower-triangular) lifting matrices with  lifting filters in $\mathscr{P}_a$ form a group, $\mathscr{U}$ (resp., $\mathscr{L}$).  In contrast to WS group lifting factorizations, concentric delay-minimized HS filter banks \emph{never} factor completely into WA lifting steps~\cite[Theorem~13]{BrisWohl06}.  The obstruction, which does not exist for WS filter banks, is the possibility that a reduced-order intermediate HS filter bank in the factorization process will correspond to filters $H_0(z)$ and $H_1(z)$ of \emph{equal lengths.}  Given  a concentric equal-length HS filter bank, it is \emph{never} possible to reduce its order  by factoring off a WA lifting step.  This leaves us with an incomplete lifting  theory for unimodular  HS filter banks.
\begin{theorem}[\cite{BrisWohl06}, Theorem~14]\label{thm_HS_factorization}
A unimodular filter bank, $\mathbf{H}(z)$, satisfies the concentric delay-minimized HS convention~(\ref{HS_anal_mirror_DS}) if and only if it can be decomposed into a partially factored lifting cascade of WA lifting steps  satisfying~(\ref{WA_lifting_filter}) and a concentric equal-length HS base filter bank $\mathbf{B}(z)$ satisfying~(\ref{HS_anal_mirror_DS}):
\begin{equation}\label{HS_lift_cascade}
\mathbf{H}(z) = \mathbf{S}_{N-1}(z) \cdots \mathbf{S}_0(z)\,\mathbf{B}(z).
\end{equation}
\end{theorem}
There is no gain-scaling matrix, $\mathbf{D}_K$, in~(\ref{HS_lift_cascade}) since $\mathbf{B}(z)$ has been left unfactored.  

One popular choice for the equal-length base  filter bank in HS lifting constructions is the Haar filter bank, which has a particularly simple lifting factorization~(\ref{Haar}). The 2-tap/10-tap HS filter bank specified in JPEG~2000 Part~2~\cite[Annex~H.4.1.1.3]{ISO_15444_2} is lifted from the Haar   via a lower-triangular 4th-order WA lifting step.
Another important example  is the 6-tap/10-tap HS filter bank  in~\cite[Annex~H.4.1.2.1]{ISO_15444_2}.  This filter bank was originally constructed by spectral factorization and  has a lifting factorization of the form
\mbox{$\mathbf{H}(z) = \mathbf{S}(z)\mathbf{B}(z)$,} 
where $S(z)$ is a second-order WA filter and $\mathbf{B}(z)$ is an equal-length (6-tap/6-tap)  HS filter bank.

Defining a class $\mathfrak{H}_r$ of \emph{reversible}  HS filter banks is  awkward; see~\cite[Example~5]{Bris:10:GLS-I}.

\section{Uniqueness of Linear Phase Lifting Factorizations}\label{sec:Unique}		
In the last section we saw that every filter bank in the unimodular WS and HS classes factors into linear phase lifting steps of an appropriate form.  Lifting factorizations, like other elementary matrix decompositions, are highly nonunique, and although  linear phase factorizations are more specialized than general lifting decompositions there seems little reason \emph{a~priori} to expect them to be unique.  There are, however, a few trivial causes of nonuniqueness that we can exclude in an \emph{ad~hoc} fashion.
\begin{definition}[\cite{Bris:10:GLS-I}, Definition~3]\label{defn:Irreducible}
A  lifting cascade~(\ref{lift_from_B}) is {\em irreducible\/} if all  lifting steps are nontrivial ($\mathbf{S}_i(z)\neq \mathbf{I}$) and there are no consecutive lifting matrices with the same update characteristic, i.e., the lifting matrices strictly alternate between lower- and upper-triangular. 
\end{definition}

Every lifting cascade can be simplified to an irreducible cascade using matrix multiplication.  Merely restricting attention to irreducible lifting cascades is far from sufficient to ensure unique factorizations, as the two irreducible lifting factorizations of the Haar filter bank  (\ref{alt_Haar_lifting_steps})--(\ref{Haar}) show.  To view nonuniqueness in a different light, move the lifting steps from~(\ref{Haar}) over to the right end of~(\ref{alt_Haar_lifting_steps}) and use~\cite[Section~7.3]{DaubSwel98} to factor $\diag(1/2,\,2)$  into lifting steps.  This results in an irreducible lifting factorization of the identity,
\begin{equation} \label{identity_lift}
\mathbf{I}=
        \left[ \begin{array}{cc}
                1 & 0 \\
               -1 & 1
        \end{array}\right]
        \left[ \begin{array}{cc}
                1 & -1 \\
                0 & 1
        \end{array}\right]
        \left[ \begin{array}{cc}
                1       & 0 \\
                1/2    & 1
        \end{array}\right]
        \left[ \begin{array}{cc}
                1   & 2 \\
                0   & 1
        \end{array}\right] 
	\left[ \begin{array}{cc}
                1  & 0 \\
                -1/2  & 1
        \end{array}\right]
        \left[ \begin{array}{cc}
                1  & 1 \\
                0  & 1
        \end{array}\right]
       \left[ \begin{array}{cc}
                1  & 0 \\
                1  & 1
        \end{array}\right]
        \left[ \begin{array}{cc}
                1  & -1/2 \\
                0  & 1
        \end{array}\right].
\end{equation}

In a similar manner, \emph{any} transfer matrix with two distinct  irreducible lifting factorizations  gives rise to an  irreducible factorization of the identity; cf.~\cite[Example~1]{Bris:10:GLS-I}, which presents an irreducible, reversible lifting factorization of the identity using linear phase (HS and HA) lifting filters.  By constructing irreducible lifting factorizations of the identity, it is possible to sharpen the universal lifting factorization result of~\cite{DaubSwel98} into the following universal \emph{nonunique} factorization result.
\begin{proposition}[\cite{Bris:10:GLS-I}, Proposition~1]\label{prop:Nonuniqueness}
If $\mathbf{G}(z)$ and $\mathbf{H}(z)$ are any FIR perfect reconstruction filter banks then $\mathbf{G}(z)$ can be irreducibly lifted from $\mathbf{H}(z)$  in infinitely many different ways.
\end{proposition}
%

\subsection{Group Lifting Structures}\label{sec:Unique:GLS}  
In light of the rich supply of elementary matrices, this plethora of irreducible lifting factorizations (almost all of which are useless for applications) results from our failure to specify precisely which liftings we regard as \emph{useful}.  The JPEG committee restricted the scope of [13, Annex G] to linear phase lifting factorizations of WS filter banks because these were considered to be the most useful liftings for conventional image coding, while [13, Annex H] was written to accommodate arbitrary lifted filter banks for niche applications.
Taking a cue from the JPEG committee, we  formalize a framework for specifying \emph{restricted universes} of lifting factorizations.  Group theory turns out to be a  convenient tool for  this task.

\subsubsection{Lifting Matrix Groups}\label{sec:Unique:GLS:Lift}
As mentioned above, upper-triangular (resp., lower-triangular) lifting matrices form multiplicative groups, $\mathscr{U}$ (resp., $\mathscr{L}$), as do lifting matrices whose  lifting filters are restricted to additive groups of  Laurent polynomials.  This includes groups of filters whose symmetry and  group delay are given, such as the groups $\mathscr{P}_0$ and  $\mathscr{P}_1$ of HS lifting filters associated with Lemma~\ref{lemma_HS_steps}.  Define abelian group isomorphisms 
\[ \upsilon,\,\lambda : \mathbb{C}[z,z^{-1}]\rightarrow\mathscr{N} \]
that map a lifting filter $S(z)\in\mathbb{C}[z,z^{-1}]$ to lifting matrices,
\begin{equation}\label{monomorphisms}
\upsilon(S) \equiv
        \left[ \begin{array}{cc}
                1  & S(z) \\
                0  & 1
        \end{array}\right]
        \quad\mbox{and}\quad
\lambda(S) \equiv
        \left[ \begin{array}{cc}
                1  & 0\\
                S(z)  & 1
        \end{array}\right].
\end{equation}
\begin{definition}[\cite{Bris:10:GLS-I}, Definition~4]\label{defn:lifting_matrix_groups}
Given two additive groups of Laurent polynomials, \mbox{$\mathscr{P}_i<\mathbb{C}[z,z^{-1}]$,} \mbox{$i=0,\,1$,}
the groups $\mathscr{U}\equiv\upsilon(\mathscr{P}_0)$ and $\mathscr{L}\equiv\lambda(\mathscr{P}_1)$ are called the \emph{lifting matrix groups} generated by $\mathscr{P}_0$ and $\mathscr{P}_1$. 
\end{definition}
\subsubsection{Gain-Scaling Automorphisms}\label{sec:Unique:GLS:Gain}
The unimodular gain-scaling matrices $\mathbf{D}_K \equiv \diag(1/K,K)$ also form an abelian  group with the product $\mathbf{D}_K \mathbf{D}_J = \mathbf{D}_{KJ}$, which says that we have an isomorphism
\begin{equation}\label{D_isomorphism}
\mathbf{D}\colon \mathbb{R}^*\equiv \mathbb{R}\backslash\{0\}\stackrel{\cong}{\longrightarrow}\mathscr{D}<\mathscr{N}. 
\end{equation}
$\mathscr{D}$ acts on $\mathscr{N}$ via inner automorphisms,
\begin{equation}\label{conjugation_operator}
\gam{K}\mathbf{A}(z) \equiv \mathbf{D}_K\,\mathbf{A}(z)\,\mathbf{D}_K^{-1} \,,\qquad
\gam{K}  \left[ \begin{array}{cc}
                    a  & b\\
                    c  & d
                    \end{array}\right]%
=
                    \left[ \begin{array}{cc}
                    a  & K^{-2} b\\
                    K^2 c  & d
                    \end{array}\right] .
\end{equation}
This is equivalent to the intertwining relation
\begin{equation}\label{intertwine_gain}
\mathbf{D}_K\,\mathbf{A}(z) = (\gam{K}\mathbf{A}(z))\,\mathbf{D} _K
\end{equation}
and makes $\gamma\colon\mathbf{D}_K\mapsto \gam{K}$  a homomorphism of $\mathscr{D}$ onto a subgroup $\gamma(\mathscr{D})<\Aut(\mathscr{N})$.  
\begin{definition}[\cite{Bris:10:GLS-I}, Definition~5]\label{defn:D_invariance}
A group $\mathscr{G<N}$ is \emph{$\mathscr{D}$-invariant} if all of the inner automorphisms \mbox{$\gam{K}\in\gamma(\mathscr{D})$} fix the group $\mathscr{G}$; i.e., \mbox{$\gam{K}\mathscr{G}=\mathscr{G}$,}  so that \mbox{$\gam{K}|_{\mathscr{G}}\in\Aut(\mathscr{G})$}.  This is equivalent to saying that
$\mathscr{D}$ lies in the normalizer of $\mathscr{G}$ in $\mathscr{N}$:
\[ 
{\mathscr{D}} \;<\; N_{\mathscr{N}}(\mathscr{G}) \;\equiv\;
\left\{\mathbf{A}\in\mathscr{N} : \mathbf{A}\mathscr{G}\mathbf{A}^{-1} = \mathscr{G}\right\}\,.
\]
\end{definition}

For instance, when the lifting filter groups $\mathscr{P}_0$ and $\mathscr{P}_1$ are \emph{vector spaces} it follows easily from~(\ref{conjugation_operator}) that $\mathscr{U}\equiv\upsilon(\mathscr{P}_0)$ and $\mathscr{L}\equiv\lambda(\mathscr{P}_1)$ are $\mathscr{D}$-invariant  matrix groups.

\subsubsection{Definition of Group Lifting Structures}\label{sec:Unique:GLS:Def}  
We now have the machinery needed to define  a ``universe'' of lifting factorizations.  In the following, $\mathfrak{B}$ denotes a set (not necessarily a group) of base filter banks from which other filter banks are lifted in partially factored lifting cascades (\ref{lift_from_B}). 
\begin{definition}[\cite{Bris:10:GLS-I}, Definitions~6 and~7]\label{defn:LiftingStructures}
A \emph{group lifting structure} is an ordered four-tuple, 
\[ \mathfrak{S}\equiv(\mathscr{D},\,\mathscr{U},\,\mathscr{L},\,\mathfrak{B}), \]
where $\mathscr{D}$ is a gain-scaling group, $\mathscr{U}$ and $\mathscr{L}$ are  upper- and lower-triangular lifting matrix groups, and $\mathfrak{B}\subset\mathscr{N}$.  
The \emph{lifting cascade group},  $\mathscr{C}$, generated by $\mathfrak{S}$ is the subgroup of $\mathscr{N}$ generated by $\mathscr{U}$ and $\mathscr{L}$:
\begin{equation}\label{lifting_cascade_group}
\mathscr{C} \equiv \langle\mathscr{U\cup L}\rangle =
\left\{\mathbf{S}_1\cdots\mathbf{S}_k \colon
k\geq 1,\;\mathbf{S}_i\in\mathscr{U\cup L}\right\}. 
\end{equation}
The \emph{scaled lifting group},  $\mathscr{S}$, generated by $\mathfrak{S}$ is the subgroup generated by $\mathscr{D}$ and $\mathscr{C}$:
\begin{equation}\label{scaled_lifting_group}
\mathscr{S}  \equiv  \langle\mathscr{D\cup C}\rangle
= \left\{\mathbf{A}_1\cdots\mathbf{A}_k \colon
k\geq 1,\;\mathbf{A}_i\in\mathscr{D\cup U\cup L}\right\}.
\end{equation}
We say  $\mathfrak{S}$ is a \emph{$\mathscr{D}$-invariant group lifting structure} if $\mathscr{U}$ and $\mathscr{L}$, and therefore $\mathscr{C}$, are $\mathscr{D}$-invariant groups.
\end{definition}

Given a group lifting structure, the universe of all filter banks generated by $\mathfrak{S}$ is 
\[  \mathscr{DC}\mathfrak{B} \equiv \left\{\mathbf{DCB} \colon
\mathbf{D}\in\mathscr{D},\;\mathbf{C}\in\mathscr{C}\,,\;\mathbf{B}\in\mathfrak{B}\right\}.  \]
The statement  ``$\mathbf{H}$ has a (group) lifting factorization in $\mathfrak{S}$''  means  $\mathbf{H}\in\mathscr{DC}\mathfrak{B}$.  $\mathbf{H}$ has a  lifting factorization in $\mathfrak{S}$ if and only if it has an \emph{irreducible}  factorization in $\mathfrak{S}$.

The group lifting structure that characterizes the universe of WS group lifting factorizations is defined as follows.  The lifting matrix groups $\mathscr{U}\equiv\upsilon(\mathscr{P}_0)$ and $\mathscr{L}\equiv\lambda(\mathscr{P}_1)$ are determined by the groups $\mathscr{P}_0$ and $\mathscr{P}_1$ of HS lifting filters defined in Section~\ref{sec:Lift:WS}.  By Theorem~\ref{thm_WS_factorization}  unimodular WS filter banks  factor completely  over $\mathscr{U}$ and $\mathscr{L}$, so we  set $\mathfrak{B} \equiv\left\{\mathbf{I}\right\}$.  Since $\mathscr{P}_0$ and $\mathscr{P}_1$ are vector spaces, setting  $\mathscr{D} \equiv \mathbf{D}(\mathbb{R}^*)$ results in a $\mathscr{D}$-invariant group lifting structure, $\mathfrak{S}_{\mathscr{W}}\equiv(\mathscr{D},\mathscr{U},\mathscr{L},\mathfrak{B})$.  The conclusion of Theorem~\ref{thm_WS_factorization} can be stated succinctly in terms of $\mathscr{C_W\equiv\langle U\cup L\rangle}$ as
\begin{equation}\label{W=DC}
\mathscr{W} = \mathscr{DC_W}\mathfrak{B} = \mathscr{DC_W}\,.
\end{equation}

The group lifting structure for  delay-minimized HS  lifting factorizations is  more complicated.  The lifting matrix groups $\mathscr{U}\equiv\upsilon(\mathscr{P}_a)$ and $\mathscr{L}\equiv\lambda(\mathscr{P}_a)$ are determined by the group $\mathscr{P}_a$  of WA lifting filters defined in Section~\ref{sec:Lift:HS}.  Per Theorem~\ref{thm_HS_factorization},  we  define $\mathfrak{B_H}$ to be the set of all concentric equal-length HS filter banks.  Defining  $\mathscr{D} \equiv \mathbf{D}(\mathbb{R}^*)$  results in a $\mathscr{D}$-invariant group lifting structure, $\mathfrak{S}_{\mathfrak{H}}\equiv(\mathscr{D},\mathscr{U},\mathscr{L},\mathfrak{B_H})$.  With $\mathscr{C}_{\mathfrak{H}}\equiv\langle U\cup L\rangle$ the conclusion of Theorem~\ref{thm_HS_factorization} can be stated  as
\begin{equation}\label{H=DCB}
\mathfrak{H} = \mathscr{DC_{\mathfrak{H}}}\mathfrak{B_H} \,.
\end{equation}
Group lifting structures $\mathfrak{S}_{\mathscr{W}_r}$ and $\mathfrak{S}_{\mathfrak{H}_r}$ for \emph{reversible} WS and HS filter banks are defined in~\cite[Section~IV]{Bris:10:GLS-I}.

\subsection{Unique Irreducible Group Lifting Factorizations}\label{sec:Unique:Unique}
We  need one more hypothesis in addition to irreducibility to infer uniqueness of group lifting factorizations within a given group lifting structure.  The key is found in the fact that nonunique lifting factorizations can be rewritten as irreducible lifting factorizations of the identity, such as~(\ref{identity_lift}).  Given a (nonconstant) lifting of the identity like~\cite[equation~(21)]{Bris:10:GLS-I}, if some partial product $\mathbf{E}^{(n)}(z)$  of lifting steps~(\ref{recursive_cascade}) has positive polyphase order  then the order of subsequent partial products must eventually \emph{decrease} because the final product, $\mathbf{I}$, has order zero.  This suggests that lifting structures that only generate ``order-increasing'' cascades will  generate \emph{unique} factorizations, an idea  that will be made rigorous in Theorem~\ref{thm:unique_factorization}.  
\begin{definition}[\cite{Bris:10:GLS-I}, Definition~10]\label{defn:OrderIncreasing}
A  lifting cascade~(\ref{lift_from_B}) is {\em strictly polyphase order-increasing\/} (usually shortened to  \emph{order-increasing}) if the order~(\ref{FB_order}) of each intermediate polyphase matrix~(\ref{recursive_cascade}) is strictly greater than that of its predecessor:
\[ 
\order\left(\mathbf{E}^{(n)}\right) > \order\left(\mathbf{E}^{(n-1)}\right)\quad\mbox{for $0\leq n < N$.}
\]
A group lifting structure, $\mathfrak{S}$, is  called order-increasing if every irreducible cascade in 
\mbox{$\mathscr{C}\mathfrak{B}$} is order-increasing.
\end{definition}
\subsubsection{An Abstract Uniqueness Theorem}\label{sec:Unique:Unique:Abstract}
\begin{theorem}[\cite{Bris:10:GLS-I}, Theorem~1]
\label{thm:unique_factorization}
Suppose that $\mathfrak{S}$
is a $\mathscr D$-invariant, order-increasing group lifting structure.  Let $\mathbf{H}(z)$ be a transfer matrix generated by $\mathfrak{S}$, and suppose we are given two irreducible group lifting factorizations of $\mathbf{H}(z)$ in \mbox{$\mathscr{D}\mathscr{C}\mathfrak{B}$}:
\begin{eqnarray}
\mathbf{H}(z) & = & \mathbf{D}_K\,\mathbf{S}_{N-1}(z) \cdots\mathbf{S}_0(z)\, \mathbf{B}(z)  
                \label{unprimed_cascade} \\
        & = & \mathbf{D}_{K'}\,\mathbf{S}'_{N'-1}(z)\cdots \mathbf{S}'_0(z)\,\mathbf{B}'(z)\;.
                \label{primed_cascade}
\end{eqnarray}
Then~(\ref{unprimed_cascade}) and~(\ref{primed_cascade}) satisfy the following three properties:
\begin{eqnarray}
N' & = & N\,, \label{NprimeEqualsN}\\
\mathbf{B}'(z) & = & \mathbf{D}_{\alpha}\,\mathbf{B}(z)\quad\mbox{where\ }\alpha\equiv K/K', \label{defn_alpha}\\
\mathbf{S}'_i(z) & = & \gamma_{\!\alpha}\mathbf{S}_i(z)\quad\mbox{for $i=0,\ldots,N-1$.} \label{almost_unique_factors}
\end{eqnarray} 
If, in addition, $\mathbf{B}(z)$ and $\mathbf{B}'(z)$ share a nonzero matrix entry at some point $z_0$ then the factorizations~(\ref{unprimed_cascade}) and~(\ref{primed_cascade}) are \emph{identical;} i.e., $K'=K$, $\mathbf{B}'(z)=\mathbf{B}(z)$, and
\begin{equation}\label{unique_factors}
\mathbf{S}'_i(z) = \mathbf{S}_i(z)\quad \mbox{for $i=0,\ldots,N-1$.}
\end{equation}
It also follows that $K'=K$ if either of the scalar base filters, $B_0(z)$ or $B_1(z)$, shares a nonzero value with its primed counterpart; e.g.,  if the base filter banks have equal lowpass DC responses.
\end{theorem}

The relationship described by (\ref{NprimeEqualsN})--(\ref{almost_unique_factors}) leads to the following definition.
\begin{definition}[\cite{Bris:10:GLS-I}, Definition~11]
\label{defn:ModuloRescaling}
Two factorizations of $\mathbf{H}(z)$ that satisfy (\ref{NprimeEqualsN})--(\ref{almost_unique_factors}) are said to be \emph{equivalent modulo rescaling.}  If \emph{all} irreducible group lifting factorizations of $\mathbf{H}(z)$ are equivalent modulo rescaling for \emph{every} $\mathbf{H}(z)$ generated by $\mathfrak{S}$, we say that irreducible factorizations in $\mathfrak{S}$ are \emph{unique modulo rescaling.} 
\end{definition}
\subsubsection{Application to WS and HS Group Lifting Structures}\label{sec:Unique:Unique:App}

Applying Theorem~\ref{thm:unique_factorization} is  nontrivial, and verifying the order-increasing property is the hardest aspect of the whole theory.  The key lemma for proving the order-increasing property for the WS and HS group lifting structures is the following result.
\begin{lemma}[\cite{Bris:10b:GLS-II}, Lemma~2]
\label{lem:SufficientConditions}
Let $\mathfrak{S}$ be a group lifting structure satisfying the following two polyphase vector conditions.
\begin{enumerate}
\item  For all $\mathbf{B}(z)\in\mathfrak{B}$, the polyphase support intervals~(\ref{poly_vector_supp_int}) for the base polyphase filter vectors are equal:
\begin{equation}\label{equal-order_base}
\suppint(\bsy{b}_0) = \suppint(\bsy{b}_1) .
\end{equation}

\item  For all irreducible lifting cascades in $\mathscr{C}\mathfrak{B}$, the  polyphase support intervals~(\ref{poly_vector_supp_int}) for the intermediate polyphase filter vectors satisfy the proper inclusions
\begin{equation}\label{support-covering}
\suppint\left(\bsy{e}^{(n)}_{1-m_n}\right) \varsubsetneq \suppint\left(\bsy{e}^{(n)}_{m_n}\right)
\quad\mbox{for $n\geq 0$.}
\end{equation}
\end{enumerate}
It then follows that $\mathfrak{S}$ is strictly polyphase order-increasing.
\end{lemma}

Hypothesis (\ref{equal-order_base}) is the correct answer to the ill-posed question, ``What do all concentric equal-length HS base filter banks have in common with the lazy wavelet filter bank, $\mathbf{I}$?''  This was one of the last pieces of the uniqueness puzzle to be solved and  unified the uniqueness  proofs for the WS and HS cases.

\begin{theorem}[\cite{Bris:10b:GLS-II}, Theorem~1]
\label{thm:WS_Uniqueness}
Let $\mathfrak{S}_{\mathscr{W}}$ and $\mathfrak{S}_{\mathscr{W}_r}$ be the group lifting structures defined in~\cite[Section~IV-A]{Bris:10:GLS-I}.
Every filter bank in $\mathscr{W}$ has a unique irreducible lifting factorization in $\mathfrak{S}_{\mathscr{W}}$ and every filter bank in $\mathscr{W}_r$ has a unique irreducible lifting factorization in $\mathfrak{S}_{\mathscr{W}_r}$.
\end{theorem}
\begin{corollary}[\cite{Bris:10b:GLS-II}, Corollary~1]
\label{cor:J2K_Pt2_Annex_G_uniqueness}
A delay-minimized unimodular WS filter bank can be specified in JPEG~2000 Part~2 Annex~G syntax in one and only one way.
\end{corollary}

The proof of Theorem~\ref{thm:WS_Uniqueness} involves deriving the support-interval covering property~(\ref{support-covering}) needed to invoke  Lemma~\ref{lem:SufficientConditions} and  Theorem~\ref{thm:unique_factorization}.  The support-interval covering property  results from the following tedious lemma based on the recursive formulation of lifting~(\ref{recursive_cascade}).  The update characteristic of $\mathbf{S}_n(z)$ (Definition~\ref{defn:update_char}) is $m_n$ and the \emph{support radius}  of a filter is the radius of its support interval,
\begin{equation}\label{supp_rad}
\supprad(f)\equiv \left\lfloor\frac{b-a+1}{2}\right\rfloor,\quad\mbox{where}\quad [a,b]=\suppint(f).
\end{equation}

%
\begin{lemma}[\cite{Bris:10b:GLS-II}, Lemma~5]
\label{lem:WS_support_formula}
Let \mbox{$\mathbf{S}_{N-1}(z) \cdots\mathbf{S}_0(z)\in\mathscr{C}_{\mathscr{W}}$} be an irreducible cascade with intermediate  scalar filters $E^{(n)}_i(z),$ $i=0,\,1.$  Let $r^{(n)}_i$ be the support radius of  $e^{(n)}_i,$ 
and let \mbox{$t^{(n)}\geq 1$} be the support radius of the HS lifting filter $S_n(z)$.
Then $\suppint\left(e^{(n)}_i\right)$ is centered at $-i$,
\[ \suppint\left(e^{(n)}_i\right) = \left[-r^{(n)}_i-i,\,r^{(n)}_i-i\right]\,,\quad i=0,\,1, \]
where
\begin{equation}\label{WS_support_radius}
r^{(n)}_{m_n} = r^{(n)}_{1-m_n} + 2t^{(n)} -1\quad\mbox{for $n\geq 0$,}
\end{equation}
\begin{equation}\label{WS_support_radius_nonlift}
r^{(n)}_{1-m_n} = r^{(n-1)}_{m_{n}} + 2t^{(n-1)} -1\quad\mbox{for $n\geq 1$,} 
\end{equation}
with $r^{(0)}_{1-m_0} = r^{(-1)}_{1-m_0}=0$.
\end{lemma}
%

There is a similar unique factorization result for unimodular HS filter banks.
\begin{theorem}[\cite{Bris:10b:GLS-II}, Theorem~2]
\label{thm:HS_Uniqueness}
Let $\mathfrak{S_H}$ and $\mathfrak{S}_{\mathfrak{H}_r}$ be the group lifting structures defined in~\cite[Section~IV-B]{Bris:10:GLS-I}.
Every filter bank in $\mathfrak{H}$ has an irreducible group lifting factorization in $\mathfrak{S_H}$ that is  unique modulo rescaling.
Every filter bank in $\mathfrak{H}_r$ has a unique irreducible group lifting factorization in $\mathfrak{S}_{\mathfrak{H}_r}$.
\end{theorem}
%



\section{Group-Theoretic Structure of Linear Phase Filter Banks}\label{sec:Group}		
We can now characterize the group-theoretic structure of the groups generated by a $\mathscr{D}$-invariant, order-increasing group lifting structure.  First we consider the lifting cascade group,  $\mathscr{C}$, which  only depends on $\mathscr{U}$ and $\mathscr{L}$, after which we consider the structure generated by scaling operations in the scaled lifting group, $\mathscr{S}$.

\subsection{Free Product Structure of  Lifting Cascade Groups}\label{sec:Group:Free}
\begin{figure}[tb]
\sidecaption
		\includegraphics{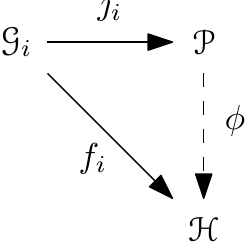}
		\caption{Commutative diagram defining a free product of the groups $\mathscr{G}_i$.}
		\label{fig:Coproduct}
\end{figure}

Recall the definition of free products in the category of groups.
\begin{definition}[\cite{Hungerford74,Rotman95}]\label{defn:Coproducts}
Let $\{\mathscr{G}_i:i\in I\}$ be an indexed family of groups, and let $\mathscr{P}$ be a group with homomorphisms 
\mbox{$\jmath_i:\mathscr{G}_i\rightarrow \mathscr{P}$.}  
Then $\mathscr{P}$ is called a \emph{free product of the groups $\mathscr{G}_i$} if and only if, for every group $\mathscr{H}$ and family of homomorphisms 
\mbox{$f_i:\mathscr{G}_i\rightarrow \mathscr{H}$,}
there exists a unique homomorphism
\mbox{$\phi:\mathscr{P}\rightarrow \mathscr{H}$}
such that $\phi\circ \jmath_i=f_i$ for all $i\in I$.  This  is equivalent to saying that there exists a unique homomorphism $\phi$ such that the diagram in Figure~\ref{fig:Coproduct} commutes for all $i\in I$.
\end{definition}

Defining free products via the universal mapping property  in Figure~\ref{fig:Coproduct} means  free products are \emph{coproducts} in the category of groups and are therefore uniquely determined (up to isomorphism) by their generators $\mathscr{G}_i$~\cite[Theorem~I.7.5]{Hungerford74},   \cite[Theorem~11.50]{Rotman95}.  There is a constructive procedure (the ``reduced word construction''~\cite{Hungerford74,Rotman95}) that generates a canonical realization of the free product of an arbitrary family of groups.  Standard notation for free products is 
\mbox{$\mathscr{P} = \mathscr{G}_1\bigast\mathscr{G}_2\bigast\cdots$.}

The intuition behind Theorem~\ref{thm:Cascade} (below) is the identification of irreducible group lifting factorizations over $\mathscr{U}$ and $\mathscr{L}$ with the group of reduced words over the alphabet $\mathscr{U}\cup\mathscr{L}$, which is the canonical realization of $\mathscr{U}\bigast\mathscr{L}$.  The reduced word construction of $\mathscr{U}\bigast\mathscr{L}$ is a somewhat technical chore when done rigorously, and it  would be a  messy affair at best  to write down and  verify an isomorphism  between the group of reduced words over $\mathscr{U}\cup\mathscr{L}$ and  a lifting cascade group in one-to-one correspondence with a collection of irreducible group lifting factorizations.  For this reason the proof presented in~\cite{Brislawn:13:Group-theoretic-structure} avoids the details of the reduced word construction and instead uses uniqueness of irreducible group lifting factorizations to show that $\mathscr{C}$ satisfies the categorical definition of a coproduct.

\subsubsection{Lifting Cascade Groups are Free Products of $\mathscr{U}$ and $\mathscr{L}$}\label{sec:Group:Free:Cascade}
An easy lemma is needed to deal with  group lifting structures whose irreducible group lifting factorizations are only unique modulo rescaling.
\begin{lemma}[\cite{Brislawn:13:Group-theoretic-structure}, Lemma~1]\label{lem:Cascade}
If $(\mathscr{D,U, L},\mathfrak B)$ is a $\mathscr{D}$-invariant, order-increasing group lifting structure with lifting cascade group $\mathscr{C\equiv\langle U\cup L\rangle}$ then irreducible group lifting factorizations in $\mathscr C$ are unique, even if irreducible group lifting factorizations of filter banks in $\mathscr{DC}\mathfrak{B}$ are only unique modulo rescaling.
\end{lemma}
\begin{figure}[tb]
\sidecaption
    \includegraphics{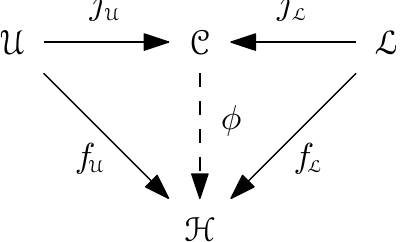}
    \caption{Universal mapping property for the coproduct $\mathscr{C\cong U \bigast L}$.}
    \label{fig:UstarL}
\end{figure}
Lemma~\ref{lem:Cascade} ensures that all $\mathscr{D}$-invariant, order-increasing group lifting structures satisfy the hypotheses of the following theorem, whose proof  consists of showing that $\mathscr{C}$ satisfies the universal mapping property  in Figure~\ref{fig:UstarL}.  
\begin{theorem}[\cite{Brislawn:13:Group-theoretic-structure}, Theorem~1]
\label{thm:Cascade}
Let $\mathscr{U}$ and $\mathscr{L}$ be upper- and lower-triangular lifting matrix groups with lifting cascade group  $\mathscr{C\equiv\langle U\cup L\rangle}$.
If every element of $\mathscr{C}$ has a unique irreducible group lifting factorization over $\mathscr{U\cup L}$ then $\mathscr{C}$ is isomorphic to the free product of $\mathscr{U}$ and $\mathscr{L}$:
\[ \mathscr{C\cong U \bigast L}\,. \]
\end{theorem}

This free product structure, $\mathscr{C\cong U \bigast L}$, is one of the conditions that are required for $\mathscr{C}$ to be a free group.
\begin{theorem}[\cite{Brislawn:13:Group-theoretic-structure}, Theorem~2]\label{thm:FreeCascadeGroup}
Let $\mathscr{C\equiv\langle U\cup L\rangle}$ be a lifting cascade group  over nontrivial lifting matrix groups  $\mathscr U$ and $\mathscr L$.   $\mathscr{C}$ is a free group (necessarily  on two generators)  if and only if $\mathscr U$ and $\mathscr L$  are infinite cyclic groups and $\mathscr{C\cong U \bigast L}$.
\end{theorem}
%

\subsection{Semidirect Product Structure of Scaled Lifting Groups}\label{sec:Group:Semi}
Consider the interaction between the gain-scaling group $\mathscr{D}$ and the lifting cascade group $\mathscr{C}$ in a scaled lifting group, $\mathscr{S\equiv\langle D\cup C\rangle}$.  As we have seen,  $\mathscr{D}$ acts on $\mathscr{C}$ via inner automorphisms so it is not surprising that, under suitable hypotheses, $\mathscr{S}$ has the structure of a semidirect product, whose definition we now review.
\begin{definition}[\cite{MacLaneBirkhoff67,Hungerford74,Rotman95}]\label{defn:SemidirectProduct}
Let $\mathscr G$ be a (multiplicative) group with identity element $1_{\mathscr G}$ and subgroups $\mathscr K$ and $\mathscr Q$.  
$\mathscr G$ is an \emph{(internal) semidirect product of $\mathscr K$ by $\mathscr Q$}, denoted $\mathscr{G=Q\ltimes K}$, if the following three axioms are satisfied. 
\begin{eqnarray}
&&\mathscr{G=\langle K\cup Q\rangle}\quad \mbox{($\mathscr K$ and $\mathscr Q$ generate $\mathscr G$)}\label{G=KvQ}\\
&&\mathscr{K\lhd G}\quad  \mbox{($\mathscr K$ is a normal subgroup of $\mathscr G$)}\label{normal_subgp}\\
&&\mathscr{K\cap Q} = 1_{\mathscr G}\quad\mbox{(the trivial group)}\label{trivial_gp}
\end{eqnarray}
\end{definition}
If $\mathscr{G=Q\ltimes K}$ then $\langle \mathscr{K\cup Q}\rangle = \mathscr{QK}$ and such product representations, $g=qk$ for $g\in \mathscr{G=QK}$, are unique.

For groups $\mathscr K$ and $\mathscr Q$ that are not subgroups of a common parent, a similar construction called an \emph{external semidirect product}, denoted $\mathscr{G=Q\ltimes_{\theta} K}$, can be performed whenever we have an automorphic group action $\theta\colon\mathscr{Q}\rightarrow\Aut(\mathscr{K})$.

\subsubsection{Scaled Lifting Groups are Semidirect Products  of $\mathscr{C}$ by $\mathscr{D}$}\label{sec:Group:Semi:Scaled}
Let $\mathfrak S=(\mathscr{D,\, U,\, L},\,\mathfrak B)$ be a  group lifting structure with lifting cascade group~$\mathscr C$ and scaled lifting group~$\mathscr S$.    The following theorem has the same hypotheses as those of Theorem~\ref{thm:unique_factorization}, but rather than invoking the unique factorization theorem the argument in~\cite{Brislawn:13:Group-theoretic-structure} proves Theorem~\ref{thm:Semidirect} directly from the hypotheses.
\begin{theorem}[\cite{Brislawn:13:Group-theoretic-structure}, Theorem~3]
\label{thm:Semidirect}
If $\mathfrak S$ is a $\mathscr D$-invariant, order-increasing group lifting structure then   $\mathscr S$ is the internal semidirect product of $\mathscr C$ by $\mathscr D$:
\[ \mathscr{S=D\ltimes C}. \]
\end{theorem}

This result can be combined with Theorem~\ref{thm:Cascade} to yield a complete group-theoretic description of the group of unimodular WS filter banks,
\[  \mathscr{W = S_W=DC_W}.  \]
\begin{corollary}[\cite{Brislawn:13:Group-theoretic-structure}, Corollary~2]
\label{cor:WS_classification}
Let $\mathfrak{S}_{\mathscr{W}}\equiv(\mathscr{D},\mathscr{U},\mathscr{L},\mathbf{I})$ be the group lifting structure for the unimodular WS  group, $\mathscr{W}$, defined in~\cite[Section~IV]{Bris:10:GLS-I}.  The group-theoretic structure of $\mathscr{W}$ is
\[  \mathscr{W} \cong\mathscr{D \ltimes_{\theta}(U\bigast L)}.  \]
\end{corollary}

A similar characterization is possible for HS filter banks.  While $\mathfrak{H}$ is not a group, the product representation
\begin{eqnarray}
\mathfrak{H} &=& \mathscr{DC_{\mathfrak{H}}} \mathfrak{B_H} = \mathscr{S}_{\mathfrak{H}} \mathfrak{B_H},\label{H_product_formula}\\
\mathfrak{B_H} &\equiv& \left\{\mathbf{B}\in\mathfrak{H}\colon\order(B_0)=\order(B_1)\right\}\label{B_HS},
\end{eqnarray}
exhibits $\mathfrak{H}$ as a collection of \emph{right cosets},  $\mathscr{S}_{\mathfrak{H}}\mathbf{B}$,  of $\mathscr{S}_{\mathfrak{H}}$ by elements of $\mathfrak{B_H}$.  These cosets do not \emph{partition} $\mathfrak{H}$, however, since they are not disjoint:  $\mathbf{B}'\equiv\mathbf{D}_{\alpha}\mathbf{B}\in\mathfrak{B_H}$ implies  $\mathscr{S}_{\mathfrak{H}}\mathbf{B}=\mathscr{S}_{\mathfrak{H}}\mathbf{B}'$.  To obtain a nonredundant partition of  $\mathfrak{H}$ into cosets, we can either eliminate scaling matrices (i.e., form cosets of $\mathscr{C}_{\mathfrak{H}}$ rather than of $\mathscr{S}_{\mathfrak{H}}$) or else normalize the elements of $\mathfrak{B_H}$.
\begin{corollary}[\cite{Brislawn:13:Group-theoretic-structure}, Corollary~3]
\label{cor:HS_classification}
Let $\mathfrak{S}_{\mathfrak{H}}\equiv(\mathscr{D},\mathscr{U},\mathscr{L},\mathfrak{B_H})$ be the group lifting structure for the unimodular HS  class, $\mathfrak{H}$, defined in~\cite[Section~IV]{Bris:10:GLS-I}.  The group-theoretic structure of $\mathscr{S}_{\mathfrak{H}}$ is
\[  \mathscr{S}_{\mathfrak{H}}\cong\mathscr{D \ltimes_{\theta}(U\bigast L)},  \]
and $\mathfrak{H}$ can be partitioned into disjoint right cosets  (but not left cosets) of either $\mathscr{C}_{\mathfrak{H}}$ or  $\mathscr{S}_{\mathfrak{H}}$:
\begin{eqnarray}	
\mathfrak{H} & = & \mbox{$\bigcup$}\left\{\mathscr{C}_{\mathfrak{H}}\mathbf{B}\colon\mathbf{B}\in\mathfrak{B_H}\right\} \label{C_cosets}\\
		& = & \mbox{$\bigcup$}\left\{\mathscr{S}_{\mathfrak{H}}\mathbf{B}\colon\mathbf{B}\in\mathfrak{B'_H}\right\}, \label{S_cosets}
\end{eqnarray}
where $\mathfrak{B'_H}$ is given by, e.g.,
\begin{equation}\label{B_HS2}
\mathfrak{B'_H}\equiv \left\{\mathbf{B}\in\mathfrak{B_H}\colon B_0(1)=1\right\}.
\end{equation}
\end{corollary}

Scaled lifting groups with the structure $\mathscr{S}\cong\mathscr{D \ltimes_{\theta}(U\bigast L)}$ have formal similarities \cite[Section~IV]{Brislawn:13:Group-theoretic-structure} to other examples in the mathematical literature of continuous groups with dilations, such as \emph{homogeneous groups} \cite{FollandStein:82:Hardy-Spaces,Stein:93:Harmonic-Analysis}.  Unlike homogeneous groups, however, scaled lifting groups are neither nilpotent nor finite-dimensional, so scaled lifting groups at present appear to be a new addition to the realm of continuous groups with scaling automorphisms.

\section{Conclusions}\label{sec:Concl}
We have surveyed recent results characterizing the group-theoretic structure of the two principal classes of two-channel linear phase perfect reconstruction unimodular filter banks, the whole-sample symmetric and the half-sample symmetric classes.  WS filter banks presented in the polyphase-with-advance representation naturally form a multiplicative  subgroup, $\mathscr{W}$, of the group of all unimodular matrix Laurent polynomials.  Although the class $\mathfrak{H}$ of unimodular HS filter banks does not form a group, lifting factorization theory shows that  HS filter banks form cosets of a particular group generated by unimodular diagonal gain-scaling matrices and  lifting matrices with whole-sample antisymmetric lifting filters.  An algebraic framework known as a group lifting structure has been introduced for formalizing the group-theoretic structure of lifting factorizations, and it has been shown that  the  group lifting structures for  WS (respectively, HS) filter banks satisfy a nontrivial polyphase order-increasing property that implies uniqueness of irreducible group lifting factorizations.  

These unique factorization results have in turn been used to characterize the structure (up to isomorphism) of the lifting cascade group and the scaled lifting group associated with each of these classes of linear phase filter banks.  Specifically, in both cases the lifting cascade group generated by the linear phase lifting matrices is the free product of the  upper- and lower-triangular lifting matrix groups, $\mathscr{C\cong U \bigast L}$.  Also in both cases, the scaled lifting group generated by the lifting cascade group and the diagonal gain-scaling matrix group has the structure of a semidirect product, $\mathscr{S}=\mathscr{DC}\cong\mathscr{D \ltimes_{\theta}(U\bigast L)}$.  In the case of WS filter banks this directly furnishes the structure of the unimodular WS group, $\mathscr{W}$, since $\mathscr{W}=\mathscr{S_W}$.  In the case of HS filter banks, $\mathfrak{H}$ is partitioned by  the family of all right cosets of $\mathscr{C}_{\mathfrak{H}}$ by concentric equal-length base HS filter banks.  Alternatively, $\mathfrak{H}$ is also partitioned by  the family of all right cosets of $\mathscr{S}_{\mathfrak{H}}$ by concentric equal-length base HS filter banks with unit lowpass DC response.

\begin{acknowledgement}
The original research  papers \cite{BrisWohl06,Bris:10:GLS-I,Bris:10b:GLS-II,Brislawn:13:Group-theoretic-structure} described in this article were supported by the Los Alamos Laboratory-Directed Research \& Development Program.  Preparation of this article was supported by the DOE Office of Science and Kristi D.\ and Reilly R.\ Brislawn.  The author also thanks the producers of the Ipe drawing editor (\texttt{http://ipe7.sourceforge.net}) and the \TeX Live/Mac\TeX\ distribution (\texttt{http://www.tug.org/mactex}).
\end{acknowledgement}


\end{document}